\documentclass[aps,prb,twocolumn,showpacs,superscriptaddress,groupedaddress]{revtex4}  
\usepackage{graphicx}
\usepackage{dcolumn}
\usepackage{bm}
\usepackage{amsmath}
\usepackage{color}
\usepackage{braket}
\usepackage{amssymb}   
\usepackage{url}
\usepackage{cleveref}
\usepackage{pdfpages}

\draft 

\begin{document}


\title{The role of disorder in NaO$_2$ and its implications for Na-O$_2$ batteries} 



\author{Oleg Sapunkov}
\affiliation{%
 Department of Mechanical Engineering, Carnegie Mellon University, Pittsburgh, Pennsylvania 15213\\
}

\author{Vikram Pande}
\affiliation{%
 Department of Mechanical Engineering, Carnegie Mellon University, Pittsburgh, Pennsylvania 15213\\
}

\author{Abhishek Khetan}
\affiliation{Institute for Combustion Technology, RWTH, Aachen, Germany, 52056}

\author{Venkatasubramanian Viswanathan}
\email{venkvis@cmu.edu}
\affiliation{%
 Department of Mechanical Engineering, Carnegie Mellon University, Pittsburgh, Pennsylvania 15213\\
}


\date{\today}

\begin{abstract}
We present a DFT study utilizing the Hubbard U correction to probe structural and magnetic disorder in $\mathrm{NaO_{2}}$, primary discharge product of Na-O$_2$ batteries. We show that $\mathrm{NaO_{2}}$ exhibits a large degree of rotational and magnetic disorder; a 3-body Ising Model is necessary to capture the subtle interplay of this disorder. MC simulations demonstrate that energetically favorable, FM phases near room temperature consist of alternating bands of orthogonally-oriented $\mathrm{O_{2}}$ dimers. We find that bulk structures are insulating, with a subset of FM structures showing a moderate gap ($<2$ eV) in one spin channel.
\end{abstract}

\maketitle 


Electrification of road transport recently led to increased interest in high-energy-density rechargeable batteries, in particular, metal-air batteries. \cite{luntz2014nonaqueous,sapunkov2015quantifying} Li-O$_2$ battery, one of the most promising, owing to its high energy density, suffers from electrolyte\cite{burke2015enhancing,abraham1996polymer,abraham1990li+,bryantsev2011predicting,bryantsev2013identification,bryantsev2012predicting,walker2013rechargeable,elia2014advanced,khetan2014solvent,burke2015enhancing} and electrode\cite{ottakam2012carbon,thotiyl2013stable,mccloskey2012twin,mccloskey2012limitations} instability, as well as limited discharge capacity.\cite{viswanathan2011electrical,aetukuri2015solvating,chen2013charging,bruce2012li,johnson2014role} Hartmann \textit{et al.} recently demonstrated a rechargeable Na-O$_2$ battery, with sodium superoxide (NaO$_2$) as the primary observed discharge product, which showed superior cycle life to state-of-the-art Li-O$_2$ batteries.\cite{hartmann2013comprehensive,hartmann2013rechargeable} Subsequently, there have been numerous theoretical and experimental efforts to understand the fundamental mechanisms that govern electrochemistry in Na-O$_2$ batteries.\cite{kang2014nanoscale,yang2015intrinsic,hartmann2013rechargeable,das2014sodium,hartmann2013comprehensive,ellis2012sodium,kundu2015emerging,kwak2015nanoconfinement,ortiz2015rate}

Accurate description of the Na-O$_2$ battery reaction mechanisms\cite{kim2013sodium,kang2014nanoscale} requires detailed understanding of electronic structure throughout the phase space of sodium oxides. The roles of nucleation,\cite{krishnamurthy2016universality,ortiz2015rate,hartmann2015discharge} nanoscale stabilization,\cite{kang2014nanoscale} and surface energetics of various sodium-oxygen compounds have been examined through a combination of density functional theory (DFT) calculations and electrochemical measurements. This has led to an improved understanding of Na-O$_2$ battery reactions, which, as proposed, constitute a combination of surface and solution mechanisms of the NaO$_2$ discharge product.\cite{adelhelm2015lithium,sapunkov2015quantifying} In certain cases, it should be noted, Na$_2$O$_2$ has been reported as the discharge product,\cite{jian2014high,araujo2015unveiling} and selectivity between NaO$_2$ and Na$_2$O$_2$ is not yet fully characterized.

Alkali superoxides are known to be highly disordered materials, both in geometric structure and magnetic ordering.\cite{smith1966antiferromagnetism,ren2013low} Presently, there is very limited understanding of geometric and magnetic disorder in NaO$_2$, due to challenges in experimentally synthesizing NaO$_2$. It is crucial to map out disorder in room-temperature NaO$_2$, considering its importance in determining electronic structure, surface energetics, and growth properties, relevant for Na-O$_2$ batteries. In this work, we aim to fill the gaps in the current understanding of the electronic structure of NaO$_2$. We perform DFT calculations incorporating the Hubbard U correction and use that to understand the geometric and magnetic disorder in NaO$_2$. We employ an Ising model accounting for magnetic and geometric degrees of freedom and find that both 2-body and 3-body nearest-neighbor interactions, which take into account both the geometric and magnetic disorder of the structure, are critical to accurately describe bulk NaO$_2$. The Ising model is used within a metropolis monte carlo framework to characterize the complete phase space as a function of temperature.  In order to determine the effect of disorder on the electronic properties of NaO$_2$, we employ hybrid density functional theory calculations and find that the calculated bandgap is strongly affected by disorder. All investigated structures exhibit a direct bandgap around 4 eV, with the exception of a small subset of ferromagnetic structures, whose direct bandgap is below 2 eV in one spin channel.

The discharge process of the Na-O$_2$ battery involves the $\mathrm{Na^{+}}$ ion coupled electron transfer reaction with O$_2$ at the cathode, which produces NaO$_2$, the primary reported discharge product.\cite{hartmann2013comprehensive,hartmann2013rechargeable} 
\begin{equation}
\mathrm{Na^{+} + O_{2} + e^{-} \rightleftharpoons {NaO_{2}}_{(s)}}
\end{equation}
A few studies do report the formation of Na$_2$O$_2$,\cite{jian2014high} which is the thermodynamically stable structure according to the Na-O phase diagram at standard temperature and pressure.\cite{kang2014nanoscale} The preferential formation of NaO$_2$ as a discharge product in room-temperature Na-O$_2$ batteries, over the thermodynamically stabler Na$_2$O$_2$, still remains a puzzle. The stable phase of bulk NaO$_2$ itself is also temperature-dependent.\cite{wriedt1987sodium,carter1953polymorphism,templeton1950crystal,kang2014nanoscale} Below 196 K, NaO$_2$ takes the $\mathrm{Pnnm}$ space group. Between 196 and 223 K, NaO$_2$ takes the Pa\={3} space group. Above 223 K, NaO$_2$ takes the Fm\={3}m space group, the primary relevant structure for room-temperature Na-O$_2$ batteries.

To explore the reaction mechanism of the Na-O$_2$ battery, we need to understand the bulk composition of NaO$_2$ and accurately describe its electronic structure and dependent properties. The oxygen dimer in NaO$_2$ behaves similar to the superoxide anion, O$_{2}^{-}$.\cite{partridge1992theoretical} Within the molecular orbital picture of ground-state (triplet) O$_2$, the 2 highest occupied orbitals are the $\mathrm{\pi_{2p}^{*}}$ antibonding orbitals, with one electron in each orbital, following Hund's Rule of Maximum Multiplicity.\cite{boyd1984quantum} The highest occupied molecular orbital of O$_{2}^{-}$ remains $\mathrm{\pi_{2p}^{*}}$, now occupied by 3 electrons, as illustrated in the appendix. This means that the O$_{2}^{-}$ dimer has $1$ unpaired electron, making O$_{2}^{-}$ magentic. The magnetic ordering of these spins is one of the key contributions to the configurational disorder of bulk NaO$_2$.

We considered bulk NaO$_2$ in the Fm\={3}m space group structure.  In this structure, sodium occupies the body centers and corners of the cubic cell, and oxygen dimers occupy face centers and edge centers of the cell, as illustrated in the appendix. We introduce the following naming scheme for these structures to describe both their geometric and magnetic arrangement. The first 4 letters, which can be A, B, C, or D, refer to the 4 possible geometric orientations of oxygen dimers. The next 4 letters, which can be P or N, refer to the net positive or negative magnetic moment of the corresponding dimers. Thus, a structure designated AAAA-PPPP has all dimers mutually parallel to each other (AAAA), and all dimers of identical, positive magnetization (PPPP), providing a ferromagnetic structure. Alternatively, a structure designated ABCD has all dimers mutually orthogonal to each other, while a structure designated PNPN or PPNN is antiferromagnetic. Formation enthalpies were calculated for 19 possible structures to explore the phase space of bulk NaO$_2$.  

Self-consistent DFT calculations were performed using the Projector Augmented Wave Method as implemented in GPAW,\cite{enkovaara2010electronic} with the Revised PBE (RPBE) exchange correlation functional.\cite{hammer1999improved} To correct electron localization in NaO$_2$, we incorporated the Hubbard U, applied on oxygen $2p$ states in NaO$_2$.\cite{hubbard1963electron,anisimov1991band,himmetoglu2014hubbard,garcia2012importance,meredig2010method} Calculations were run with a real-space grid of 0.18 {\AA}, and 6$\times$6$\times$6 k-point sampling, following the Monkhorst-Pack scheme.\cite{monkhorst1976special} Fermi-Dirac smearing of 0.01 eV was used to facilitate convergence and broyden-type mixing of electron densities was used in the calculation.\cite{johnson1988modified,srivastava1984broyden}

Two schemes were used to calculate formation enthalpy of bulk NaO$_2$. The formation enthalpy of NaO$_2$ is given by:
\begin{equation}
\mathrm{{\Delta}H_{F_{NaO_{2}}} = E_{NaO_{2}}^{DFT} - E_{Na}^{Ref} - E_{{O_2}}^{Ref} + \Delta pV}
\end{equation}
The pressure-volume work term, $\mathrm{\Delta pV}$, can be disregarded, as it is typically about 5 orders of magnitude smaller than internal energy contributions in formation enthalpy calculations.\cite{aydinol1997ab,obrovac2014alloy} Internal energies of Na, O$_2$, and NaO$_2$ were calculated using DFT simulations.

It is well known that molecular oxygen is poorly described in DFT.\cite{droghetti2008predicting} The energy of O$_2$ can be more accurately calculated using water as the reference molecule.\cite{rossmeisl2005electrolysis} In the first scheme used, reference energy of O$_2$ was computed using the DFT-calculated internal energies of H$_2$O and gaseous H$_2$, as well as the experimental formation enthalpy of H$_2$O:
\begin{equation}
\mathrm{E_{O_2}^{Ref} = 2E_{H_{2}O}^{DFT} - 2E_{H_{2}}^{DFT} - {\Delta}H_{H_{2}O}^{Exp}}
\end{equation}
The Hubbard U correction was applied on the oxygen atoms in both H$_2$O and NaO$_2$.

The second scheme was used to calculate reference energy of bulk Na. It was demonstrated in prior work that formation enthalpies of alkali oxides, peroxides and superoxides are best described when the oxidation state of the metal in the reference compound is matched to the oxidation state of the metal in the compound under investigation.\cite{christensen2015reducing} Following this scheme, the Na reference energy was calculated using the calculated internal energies of NaCl and gaseous Cl$_2$, as well as the experimental formation enthalpy of NaCl:
\begin{equation}
\mathrm{E_Na^{Ref} = E_NaCl^{DFT} - \frac{1}{2}E_{{Cl}_2}^{DFT} - {\Delta}H_NaCl^{Exp}}
\end{equation}

Currently, there is no universally accepted method for selecting the optimal Hubbard U to fit calculations to experimental data. In this study, we follow the scheme that compares the calculated formation enthalpy of a substance under investigation to the experimentally measured formation enthalpy.\cite{wang2006oxidation} The experimentally measured formation enthalpy of NaO$_2$ is $\mathrm{{\Delta}_{f}H^{0}_{solid} = -2.7 eV}$.\cite{chase1998nist} 

We find that the formation enthalpies of all considered bulk configurations of NaO$_2$ lie within 0.3-0.6 eV/NaO$_2$, at all investigated values of the Hubbard U, as shown in Fig. 1(a). The fully-organized, ferromagnetic (AAAA-PPPP) and fully disorganized, antiferromagnetic (ABCD-PPNN) structures were found to be least energetically stable, while the moderately disorganized structures with intermediate magnetic order, such as AABB-PPPN, were most stable. Many of these moderately disorganized structures were energetically degenerate at all examined values of the Hubbard U. We observe that U = 3 eV is matches the experimental formation energy and will be used for further analysis. It is also worth highlighting that the conclusions regarding disorder in bulk NaO$_2$ remain consistent for all examined values of the Hubbard U.

Given the high degeneracy observed among examined structures of NaO$_2$, it becomes crucial to examine the role of thermal disorder and its role in determining electronic properties of NaO$_2$.  The computation of disorder in larger supercells directly through DFT is computationally intractable.  We employ the approach of determining a lattice Hamiltonian\cite{hamer1981finite} that can be used to describe the energetics of bulk NaO$_2$.

In order to map out energetic interactions between the magnetic and geometric degrees of freedom, we utilize a modified Ising Model for the lattice Hamiltonian. As implemented, the model consists of a lattice of $N$ sites $i$, whose filling is described by occupation terms, $\sigma_{i}$. System energy contributions due to particle interactions in neighboring sites are captured by the 2-body and 3-body interaction terms, $j_{2_{i,k}}$ and $j_{3_{i,k,l}}$. The total energy of the $N$-site lattice is calculated as:
\begin{equation}
\mathrm{E = -\sum_{\braket{ik}} j_{2_{i,k}}\sigma_{i}\sigma_{k} -\sum_{\braket{ikl}} j_{3_{i,k,l}}\sigma_{i}\sigma_{k}\sigma_{l}}
\end{equation}

Formation enthalpies of $\mathrm{NaO_{2}}$ calculated by DFT simulations were used to fit Ising Model coefficients for our system. Formation enthalpy data at each value of the Hubbard U was used to derive a set of corresponding interaction coefficients, though a least-squares regression fit. The sole differences among the structures lay in the relative geometric and magnetic arrangement of the oxygen dimers, and these differences are reflected in the nearest-neighbor interaction terms $j$.
A total of 4 different types of $j_{2_{i,k}}$ terms are identified, along with 6 different types of $j_{3_{i,k,l}}$ terms. Our naming scheme relies on comparison of dimers, e.g. $j_{2AAPP}$, corresponds to the interaction between 2 dimers of parallel geometry and identical magnetization, while $j_{2ABPN}$, corresponds to the interaction between 2 dimers of orthogonal geometry and opposite magnetization. Likewise, $j_{3AAAPPP}$, corresponds to the interaction among 3 dimers of mutually parallel orientation and mutually identical magnetic moment, while $j_{3ABCPPN}$ corresponds to the interaction among 3 oxygen dimers of mutually orthogonal orientation, where 2 dimers share an identical magnetization that differs from that of the third dimer. The plot of fitted interaction coefficients is shown in Fig. 1(b). The plot demonstrates that among the 2-body interaction terms, $j_{2AAPN}$ was most energetically stabilizing, while among the 3-body interaction terms, $j_{3AABPPP}$ was most energetically stabilizing. 

Our DFT calculations suggested that the most thermodynamically stable configurations correspond to bulk structures of intermediate disorder, both in geometric and magnetic arrangement (e.g. AABB-PPPN). Subsequent Ising Model calculations supported this observation, with $j_{2AAPP}$ the most-stabilizing 2-body interaction term, and $j_{3AABPPP}$ the most-stabilizing 3-body interaction term. It should be noted that using a purely 2-body Ising Model of NaO$_2$ finds that the most-stabilizing nearest-neighbor interaction term was $j_{2ABPN}$, implying that the fully-disorganized antiferromagnetic structure (e.g. ABCD-PNPN) would be the predominant energetically stable structure. This is not consistent with our DFT calculations, and therefore shows the need to incorporate 3-body interaction terms into the model.

Interaction coefficients $j$ derived from the Ising Model were used in Metropolis-Hastings Monte Carlo (MHMC) simulations, to characterize larger bulk cells with higher degree of disorder, at finite temperature. Supercell structures were constructed as N$\times$N$\times$N arrays of cubic unit cells of $\mathrm{Na_{4}O_{8}}$, the structures studied in DFT simulations. Simulations were run for supercells of size 4 to 7, to study convergence of bulk properties as the overall simulated system size increases. Periodic boundary conditions in all directions were implemented and the systems were studied using the Markov Chain Monte Carlo Method,\cite{gilks2005markov,hastings1970monte} modified with the Metropolis-Hastings Algorithm.\cite{chib1995understanding} Interaction coefficients corresponding to a Hubbard U correction of 3 eV were used with the MHMC simulations, to most closely match the experimental formation enthalpy of NaO$_2$. The results would remain qualitatively consistent for other values of U, and there would be a minor change in the observed phase transition temperatures.

At the initialization of each MHMC simulation, a fully-organized supercell bulk structure was set up, with all dimers identical (e.g. AAAA-PPPP). Each individual trial step consisted of switching a randomly selected $\mathrm{O_{2}}$ dimer in the bulk structure to one of the other available configurations; orientational or magnetic change.   The effect of this switch on system energy was calculated. The standard Metropolis Algorithm was used to decide whether to accept the trial step or not. At the end of every step, system formation enthalpy and entropy were recorded for further analysis.

Configurational entropy was calculated as the logarithm of the number of configurations with the same proportions of oxygen dimers of different types. If there are $M$ distinct dimer types available, the total number of possible configurations, $\Omega$, of the full structure can be calculated using the complete multinomial coefficient:
\begin{equation}
\Omega = \prod_{j=1}^{M} {{\sum_{i=j}^{M} N_{i}} \choose N_{j} }
\end{equation}
where $i$, $j$ refer to the available configurations of dimers and $N_{i}$, $N_{j}$ refer to the number of dimers of a particular configuration present in the supercell. The configurational entropy is then simply calculated by Boltzmann's Entropy Formula: $S_{Conf} = k_{B} \log{\Omega}$.

To explore high-temperature, high-energy phases of bulk $\mathrm{NaO_{2}}$, each system simulation was initially raised to a temperature of 1252 K.
Once the energy of the structure was stabilized at the initial high temperature, the system was annealed\cite{vanderbilt1984monte} in temperature steps of 0.25 K, down to 2 K. At each 50 K increment, formation enthalpy and configurational entropy data was collected for analysis. This scheme worked robustly and the thermodynamic quantities converged for all supercell sizes investigated.

MHMC simulations were used to determine effect of temperature on the structural disorder and in turn the free energy of bulk $\mathrm{NaO_{2}}$, as shown in Fig. 1(c) and Fig. 1(d). When the temperature is high, most of the structures could have become accessible, as the entropic term dominates. 
Thus, the structure at high temperature was predominantly fully-disordered, both in geometry and in magnetic moment as shown in Fig. 1(e). Around 650 K, the structure underwent a phase change, wherein the bulk $\mathrm{NaO_{2}}$ formed large, parallel planes of alternating geometry of $\mathrm{O_{2}}$ dimers, as depicted in Fig. 1(f). In this phase, the magnetic moment was still largely disordered, within and across the monolithic-geometry planes. Around 350 K, the bulk structure underwent a second phase change, where the magnetic moment of all dimers was locked in phase together and the geometry maintained the alternating planar structure. The degree of disorder in the system could strongly affect the electronic properties of NaO$_2$, so we explored this further using hybrid density functional theory calculations.

\begin{figure*}[!htb]
 \centering
 \includegraphics[width=\textwidth]{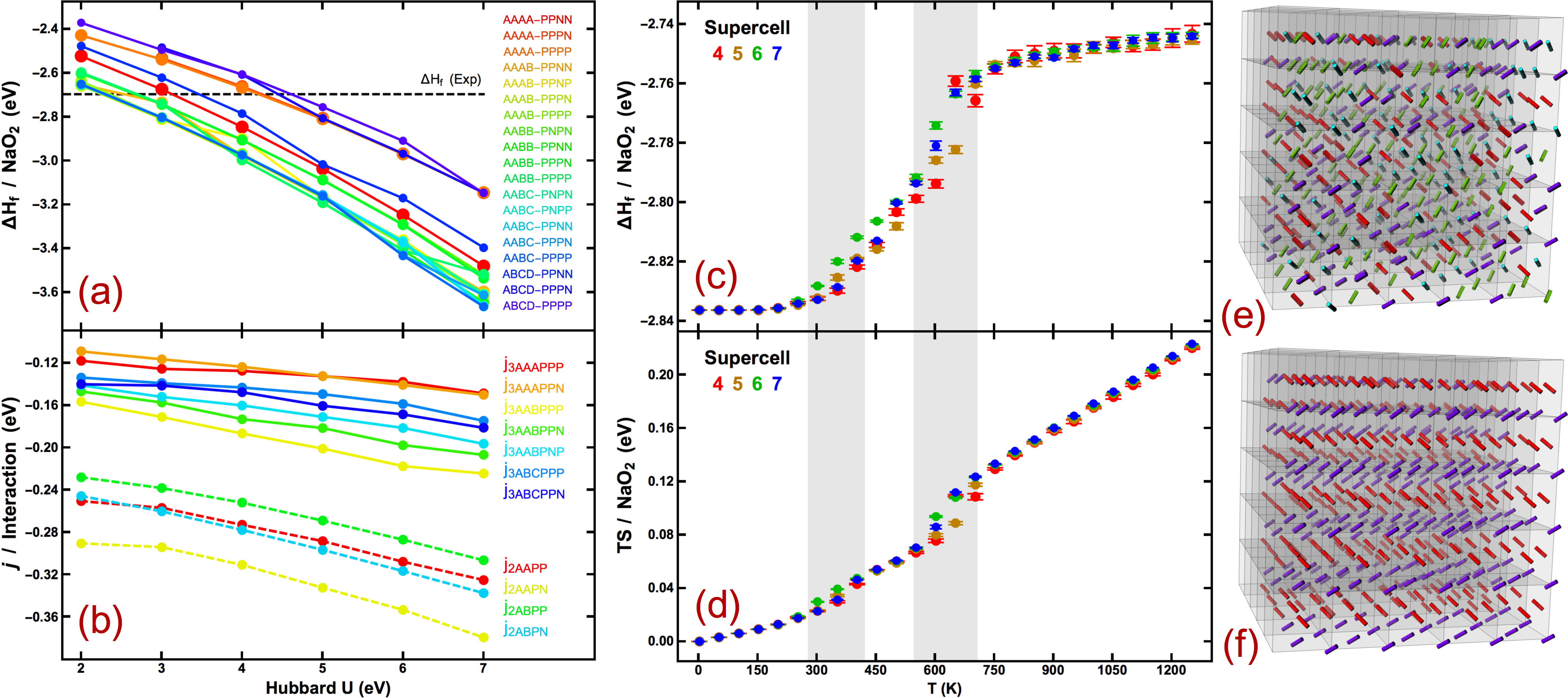}
\caption{
(a) Formation enthalpies of NaO$_2$ bulk structures were calculated for 19 4-formula-unit unit cells, using the H$_{2}$O-NaCl correction scheme, at all examined values of Hubbard U. 
~ (b) Calculated Ising Model nearest-neighbor interaction terms $j$ as function of U, per interaction within the unit cell. 
 The 3-body interactions play an important contribution. (c, d) Monte Carlo simulations using Ising model for the formation enthalpy and entropy (TS). (e) Disordered structures observed at high temperature above 700 K and (f) Geometrically aligned structures observed at around 250 K from the Monte Carlo simulations.}
\end{figure*}

Non-self-consistent hybrid density functional theory calculating employing the HSE06\cite{heyd2003hybrid} functional was used to determine direct bandgap for all phases of NaO$_2$, as an indicator of electrical conductivity. HSE06 has been shown to accurately capture bandgaps for various semiconducting and insulating materials with mean absolute error (MAE) varying from 0.26$-$0.4 eV for different classes.\cite{heyd2005energy,tran2009accurate,henderson2011accurate,chan2010efficient}  Bandgap values were calculated across all k-points and the minimum was reported for each phase of NaO$_2$ for Hubbard U = 3-7 eV.   We find that there is an increase in bandgap with increasing U as shown in Table II in S.I. The widening of the bandgap with increasing U is due to enhanced localization of electrons in the $\pi^*_{2p}$ oxygen bands in NaO$_2$. We also observe that the bandgap in both spin channels is identical for antiferromagnetic structures, and markedly different for ferromagnetic structures. This behavior is observed as the additional electrons in O$_{2}^{-}$ ions only occupy one spin channel. In the AFM structures, electrons are equally distributed in the spin channels, and the gap is between pairs of $\pi^*_{2p}$ states. In the FM structures, the bandgap in the occupied spin channel is between $\pi^*_{2p}$ and $\sigma^*_{2p}$ state and is thus higher than AFM; while for the other channel, it is between deeper $\pi^*_{2p}$ states and is thus lower than that in AFM. For mixed cases (e.g. PPPN), we expect the bandgap to be bounded by the AFM and FM cases, due to electron occupation of the $\pi^*_{2p}$ bands. Finally, we observe that the most symmetric structures, AAAA-PPPP and ABCD-PPPP. 

\begin{table}[!htb]
\begin{tabular}{|c|c|c|c|c|c|c|}
\hline
Structure & Spin 1(+) & Spin 2(-) \\
\hline
AAAA(PPPP)	&6.65 & 1.26\\
\hline
AAAB(PPPP)	&7.43 & 3.69\\
\hline
AABB(PPPP)	&7.03 & 3.96\\
\hline
AABC(PPPP)	&7.3 & 3.92\\
\hline
ABCD(PPPP)	&5.91 & 1.32\\
\hline
AAAA(PPNN)	&3.82 & 3.82\\
\hline
AAAB(PPNN)	&3.91 & 3.91\\
\hline
AABB(PPNN)	&4.13 & 4.13\\
\hline
AABB(PNPN)	&4.21 & 4.21\\
\hline
AABC(PPNN) &	4.39 & 4.13\\
\hline
ABCD(PPNN)	&3.94 & 3.94 \\
\hline
\end{tabular}
\caption{Calculaed bandgap values (eV) for various ferromagnetic and antiferromagnetic configurations of NaO$_2$, for U=3 eV using HSE06 functional.}
\label{bg}
\end{table}

In conclusion, we demonstrate the need and effectiveness of a 2-body and 3-body nearest-neighbor Ising Model in describing energetics of bulk NaO$_2$ across a large temperature range. Our study demonstrates that at low temperatures, close to room temperature and below, the predominant phase of NaO$_2$ is ferromagnetic, with alternating planes of oxygen dimers in consistent geometric orientations. The system is expected to exhibit some degree of magnetic disorder even at room temperature.  Our study on the electronic properties shows that NaO$_2$ is a wide-bandgap insulator, with a bandgap around 4 eV and we expect it to have poor electrical conductivity at room-temperature. In the context of Na-O$_2$ batteries, this implies that growth of the discharge product is most likely occurring due to the solution mechanism pathway involving a chemical dissolution of NaO$_2$ into Na$^+$ and O$_2^-$, similar to what is seen in Li-O$_2$ batteries.  However, our analysis shows that preferential nucleation of certain magnetic phases could be possible through appropriate electrode choice.  Further, our approach is expected to be important for other metal-oxygen batteries, such as K-O$_2$ batteries.

Acknowledgment is made to the Donors of the American Chemical Society Petroleum Research Fund and National Science Foundation CAREER award CBET-1554273 for partial support of this research.

\bibliography{PD_PB_Bibl}



\section*{Supplementary material}

# Supplemental Information for: The role of disorder in NaO$_2$ and its implications for Na-O$_2$ batteries

Oleg Sapunkov,[1] Vikram Pande,[1] Abhishek Khetan,[2] and Venkatasubramanian Viswanathan[1, *]

[1]*Department of Mechanical Engineering, Carnegie Mellon University, Pittsburgh, Pennsylvania 15213*
[2]*Institute for Combustion Technology, RWTH, Aachen, Germany, 52056*

(Dated: February 17, 2017)

## I. SUPPLEMENTAL INFORMATION

Self-consistent Density Functional Theory (DFT) calculations were performed using the Projector Augmented Wave (PAW) Method as implemented in GPAW.[1] The energy calculations used the Revised PerdewBurkeErnzerhof (RPBE) exchange correlation functional.[2] To correct for electron localization in NaO$_2$, we incorporated onsite electron repulsion using the Hubbard model, with the Hubbard U applied on the oxygen 2p states in NaO$_2$. For all investigated structures, the Hubbard U correction was ramped in increments of 0.1 eV.[3–7] All calculations were run with a real-space grid of 0.18 Å spacing, and a 6×6×6 k-point sampling, following the Monkhorst-Pack scheme.[8] Fermi-Dirac smearing of 0.01 eV was used to facilitate convergence and ensure accuracy. The Poisson equation was solved with convergence tolerance of $\epsilon = 10^{-12}$, to ensure electron density is solved precisely, to improve convergence of the calculations. Broyden-type mixing of electron densities was used in the calculation, mixing 5 previous densities with weights of 5%.[9] The spin densities were mixed separately[10] and the calculations were converged to a force of < 0.01 eV/Å.

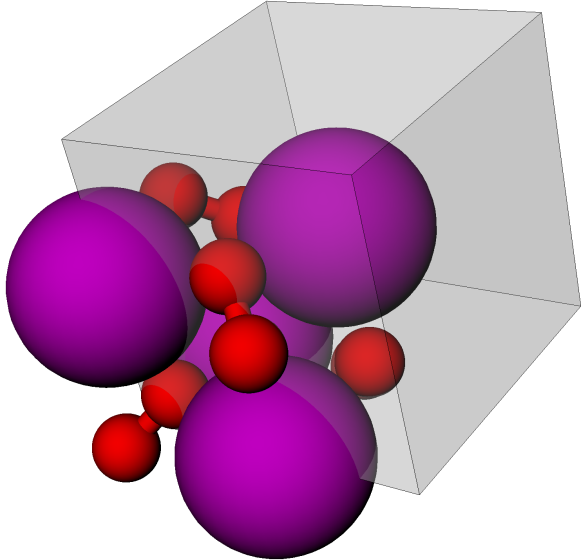

FIG. 1: Unit cell of Fm$\bar{3}$m NaO$_2$ used in the study.

For calculations involving the electronic properties of NaO$_2$, we used the HSE06 hybrid functional,[11] which has been demonstrated to give accurate bandgap estimates for a broad range of materials.[12] HSE06 is implemented in plane-wave (PW) mode in GPAW, so all structures were re-converged to the same accuracy as in the grid mode mentioned above. PW-mode calculations were first run using PBE, then the HSE06 hybrid functional was applied to correct for the energy levels of various bands. The bandgap was converged with respect to the k-point sampling, PW energy cutoff, and the Fermi-Dirac smearing. After checking convergence, k-point sampling of 8×8×8, PW energy cutoff of 600 eV, and Fermi-Dirac smearing of 0.1 eV were used for fast and accurate convergence. Input structures to the HSE06 bandgap calculations were run using the Hubbard U correction, and bandgap variation was evaluated as a function of U.

The enthalpy of formation $\Delta H_F$ of a molecule is the enthalpy required to assemble the molecule from individual atoms. $\Delta H_F$ is computed as the difference between DFT-calculated internal energy of the assembled molecule and the sum of reference internal energies of constituent species. Thus, the formation enthalpy of NaO$_2$ was calculated

using reference sodium and oxygen energies, given by:

$$\Delta H_{F_{NaO_2}} = E_{NaO_{2\,DFT}} - E_{Na_{Ref}} - E_{O_{2\,Ref}} + \Delta pV. \quad (1)$$

The pressure-volume work term, $\Delta pV$, can be disregarded, as it was found to be 5 orders of magnitude lesser than internal energy contributions in formation enthalpy calculations.[13,14] Internal energies for Na, $O_2$, and $NaO_2$ were calculated using DFT simulations. Several schemes were used to calculate reference energies for sodium and oxygen, as illustrated in Fig. 3.

Initial calculations were run using direct DFT calculations for both the bulk Na reference internal energy and the gaseous $O_2$ reference internal energy for $NaO_2$ directly. This scheme is sufficient for the calculation of formation enthalpies of most compounds.

$$E_{Na_{Ref}} = E_{Na_{U_{DFT}}} \quad (2)$$

$$E_{O_{2\,Ref}} = E_{O_{2\,U_{DFT}}} \quad (3)$$

The scheme used the ground state of molecular oxygen: triplet oxygen, which has a non-zero net magnetic moment, since both of its highest-energy electrons, in the 2p $\pi^*$ orbital, are spin-up electrons, occupying distinct orbitals, following the Pauli Exclusion Principle. The calculation was carried out with spin polarization enabled, with the two oxygen atoms preset to the same net magnetic moment of $+1/2$. In this configuration, the energy of the oxygen was over-stabilized, and the calculated formation enthalpy was too high for all values of the Hubbard U, as compared to the experimental formation enthalpy.

It is well-known that molecular oxygen is poorly described in DFT, and our first two calculation schemes for the formation enthalpy of $NaO_2$ were unable to match the experimental formation enthalpy using the DFT-calculated energy of $O_2$ alone. The energy of oxygen can be more accurately calculated using water as the reference.[15,16] In the first scheme used, reference energy of oxygen was computed using the DFT-calculated internal energies of water and gaseous hydrogen, as well as the experimental formation enthalpy of water. This correction is given by:

$$E_{O_{2\,Ref}} = 2E_{H_2O_{DFT}} - 2E_{H_{2\,DFT}} - \Delta H_{H_2O_{Exp}} \quad (4)$$

DFT-calculated sodium internal energy was used as the sodium reference in this scheme directly, i.e. $E_{Na_{Ref}} = E_{Na_{DFT}}$. The Hubbard U correction was applied on the oxygen atoms in $H_2O$, to appropriately calculate formation enthalpy of $NaO_2$ at corresponding values of the Hubbard U. Using this reference energy scheme, we were able to match the formation enthalpy of $NaO_2$, at a Hubbard U value of approximately 5.5 eV.

We also analyzed a second reference scheme, to correct for the reference energy of bulk Na. It was demonstrated in prior work that formation enthalpies of alkali oxides, peroxides and superoxides are best described when the oxidation state of the metal in the reference compound is matched to the oxidation state of the metal in the compound under investigation.[17,18] Following this scheme, the Na reference energy was calculated using the simulated internal energies of NaCl and gaseous $Cl_2$, as well as the experimental formation enthalpy of NaCl. The correction is given by:

$$E_{Na_{Ref}} = E_{NaCl_{DFT}} - \frac{1}{2}E_{Cl_{2\,DFT}} - \Delta H_{NaCl_{Exp}} \quad (5)$$

This scheme also made use of the water reference for oxygen energy, as described above. It was found that using both schemes together decreased the Hubbard U required to match the computed formation enthalpy of $NaO_2$ to its experimental value, as compared to using the water scheme alone, from approximately 5.5 eV to just over 3 eV, as shown in Fig. 3.

In order to map out the energetic interactions between the magnetic and rotational degrees of freedom, we utilize a modified Ising model. The Ising Model was orginally developed to study properties of interacting lattice systems, such as ferromagnetic materials.[19] In the model, an arbitrary lattice of $N$ sites is set up. A given site $i$ can be filled with a particle with some relevant property specified, such as spin, and is assigned an occupation term, $\sigma_i$, where $\sigma_i = 0$ if site $i$ is empty, and $\sigma_i = 1$ if site $i$ is occupied. The energy contribution due to the presence of a particle in site $i$ is designated as the field term, $h_i$. Energy contributions due to particle interactions in neighboring sites are captured by the nearest-neighbor interaction term, $j_{i,k}$ where $i$ and $k$ are two distinct sites. These nearest-neighbor interactions can be attractive or repulsive. Further interactions can also be accounted for, such as next-nearest-neighbor interactions. These would be assigned a different set of interaction terms $j_{i,k}$. The total energy of the $N$-site lattice is then calculated as:

$$E = -\sum_{i=1}^{N} h_i \sigma_i - \sum_{\langle ij \rangle} j_{i,k} \sigma_i \sigma_k \quad (6)$$

The Ising Model allows derivation of a reduced-order Hamiltonian for the system under consideration. Simulated formation enthapies are used to calculate relevant Ising Model coefficients for the system. Through the use of coefficients derived with the Ising Model, the new Hamiltonian can be implemented in a Monte Carlo simulation of larger bulk structures. These simulations introduce and allow the quantification of the effect of thermal disorder on the system.

Coefficients derived from the Ising Model were then used to simulate larger bulk cells with higher degree of disorder, to predict their formation enthalpies. In this model, we accounted for periodic boundary conditions in all directions to make our structure characteristic of a bulk material.

Systems were studied using the Markov Chain Monte Carlo Method,[20,21] modified with the Metropolis-Hastings Algorithm.[22] Since our simulated $NaO_2$ structures only included nearest-neighbor interactions, and no longer-range interactions, our Ising Model analysis likewise was limited only to nearest-neighbor interactions. We accounted for both two-body and three-body nearest-neighbor interactions. Derived nearest-neighbor $j_{2:i,k}$ and $j_{3:i,k,l}$ were used to calculate the full system formation enthalpy through the Ising model for these larger, bulk-like systems.

At the beginning of the Monte Carlo simulation, a fully-organized initial supercell structure was set up. Each individual trial step taken in the Monte Carlo simulation consisted of switching a randomly selected $O_2$ dimer in the bulk $NaO_2$ structure to one of the other available configurations for that dimer. Before and after the switch, full system formation enthalpy was calculated using the 2- and 3-body Ising Model coefficients derived earlier. To make the calculation efficient, only nearest-neighbor interactions around the altered dimer itself were calculated, since the remainder of the system maintained the same net enthalpy before and after the trial step executed. The Metropolis Algorithm was used to decide whether to accept the trial step or not:

1. A uniformly distributed random real number $\epsilon$ was chosen, between 0 and 1

2. A coefficient, called here $\beta$, was calculated: $\beta = e^{(-1/k_B T)*(E_{trial}-E_{ref})}$

3. The two numbers were compared

4. If $\epsilon \leq \beta$, the trial step was accepted, the new structure was saved as the reference structure for the next step.

5. If instead $\epsilon > \beta$, the trial step was rejected and the system was reverted back to the earlier, reference structure.

At the end of every trial step, the system formation enthalpy and entropy were recorded for further analysis.

To explore high-temperature, high-energy phases of bulk $NaO_2$, each simulation was initialized with a fully-organized supercell structure, composed of N x N x N cells of configuration AAAA-PPPP. Initial system temperature used in the Metropolis-Hastings scheme was raised to 1252 K, to examine if the simulated bulk material would maintain long-range orientational or magnetic order at high temperature. The structure was simulated at 1252 Kelvin for a number of steps $N_{st}$ sufficient to reach and maintain stable system energy. The required $N_{st}$ was scaled with the supercell size N, as $N_{st} = 15000 * (N/3)^2$. Once the energy of the structure was stabilized at the initial high temperature, the system was annealed[23] in temperature steps of 0.25 K, down to 2 K. At each 50 K increment, the structure was held for an equivalent $N_{st}$, and both formation enthalpy and configurational entropy data was collected for analysis. This scheme worked robustly for the system studied, which contained between 256 and 1372 sites. Both entropy and formation enthalpy, per oxygen dimer contained, remained consistent for all supercell sizes investigated.

The simulation was used to output data regarding the formation enthalpy of the bulk supercell, per formula unit, and the energy contribution by the configurational entropy of the bulk supercell (T ∗ S), per formula unit, as recorded at the end of every trial step. Configurational entropy was calculated as the logarithm of the number of configurations with the same proportions of oxygen dimers of different types. If there are $M$ distinct dimer types available, the total number of possible configurations, $\Omega$, of the full structure can be calculated using the complete multinomial coefficient:

$$\Omega = \prod_{j=1}^{M} \binom{\sum_{i=j}^{M} N_i}{N_j} \qquad (7)$$

where $i$, $j$ refer to the available configurations of dimers and $N_i$, $N_j$ refer to the number of dimers of a particular configuration present in the supercell. The configurational entropy is then simply calculated by Boltzmann's Entropy Formula: $S_{Conf} = k_B \log \Omega$.

| Hubbard U  | 2      | 3      | 4      | 5      | 6      | 7      |
|------------|--------|--------|--------|--------|--------|--------|
| AAAA-PPNN  | -2.523 | -2.675 | -2.846 | -3.036 | -3.248 | -3.480 |
| AAAA-PPPN  |        | -2.533 | -2.662 | -2.806 | -2.968 | -3.146 |
| AAAA-PPPP  | -2.428 | -2.538 | -2.668 | -2.810 | -2.971 | -3.145 |
| AAAB-PPNN  | -2.653 | -2.736 | -2.977 | -3.169 | -3.382 | -3.598 |
| AAAB-PPNP  | -2.658 | -2.810 | -2.900 | -3.168 | -3.362 | -3.609 |
| AAAB-PPPN  | -2.658 | -2.810 | -2.980 | -3.164 | -3.382 | -3.611 |
| AAAB-PPPP  | -2.638 | -2.804 | -2.969 | -3.171 | -3.383 | -3.616 |
| AABB-PNPN  | -2.603 | -2.744 | -2.906 | -3.088 | -3.290 | -3.521 |
| AABB-PPNN  | -2.600 | -2.740 | -2.996 | -3.192 | -3.409 | -3.646 |
| AABB-PPPN  | -2.600 | -2.742 | -2.906 | -3.089 | -3.293 | -3.536 |
| AABB-PPPP  | -2.598 | -2.740 | -2.996 | -3.192 | -3.409 | -3.517 |
| AABC-PNPN  | -2.655 | -2.802 | -2.974 | -3.160 | -3.375 | -3.607 |
| AABC-PNPP  | -2.654 | -2.804 | -2.973 | -3.162 | -3.366 | -3.662 |
| AABC-PPNN  | -2.655 | -2.803 | -2.973 | -3.161 | -3.372 | -3.666 |
| AABC-PPPN  | -2.651 | -2.802 | -2.973 | -3.156 | -3.431 | -3.614 |
| AABC-PPPP  | -2.652 | -2.804 | -2.974 | -3.162 | -3.433 | -3.665 |
| ABCD-PPNN  | -2.478 | -2.621 | -2.785 | -3.017 | -3.171 | -3.396 |
| ABCD-PPPN  |        | -2.484 | -2.608 | -2.806 | -2.968 | -3.146 |
| ABCD-PPPP  | -2.371 | -2.494 | -2.606 | -2.755 | -2.910 | -3.148 |

TABLE I: Calculated formation enthalpies of NaO$_2$ bulk structures, in $eV$. AAAA-PPPN and ABCD-PPPN structures could not be converged at a Hubbard U value of 2 eV since it did not localize the highest-energy electrons on the oxygen dimers sufficiently to stabilize the structure.

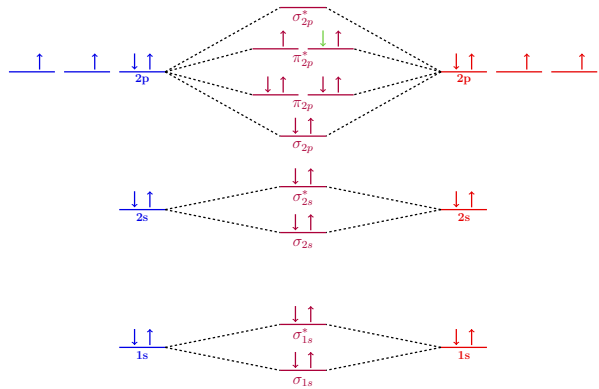

FIG. 2: Molecular Orbital Diagram of O$_2^-$. The external electron added to the O$_2$ molecule falls into the $\pi_{(}^{*}2p)$ orbital, already occupied by two spin-up electrons. This leaves oxygen with only one unpaired electron, which makes it weakly paramagnetic.

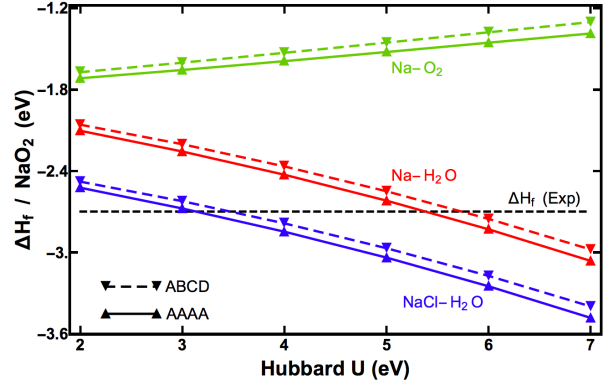

FIG. 3: Performance of different schemes used for Formation Enthalpy calculations. The simulations done with oxygen energy calculations to provide the $O_2$ reference energy gave the poorest results, since oxygen is poorly handled in DFT simulations. Correcting the oxygen energy with the water scheme improved the energy calculations tremendously, and allowed to match experimental formation enthalpy at a Hubbard U value around 6 eV. The required value of the Hubbard U was decreased to 3 eV by applying the scheme which accounted for the correct sodium ion oxidation state, using NaCl as the reference calculation.

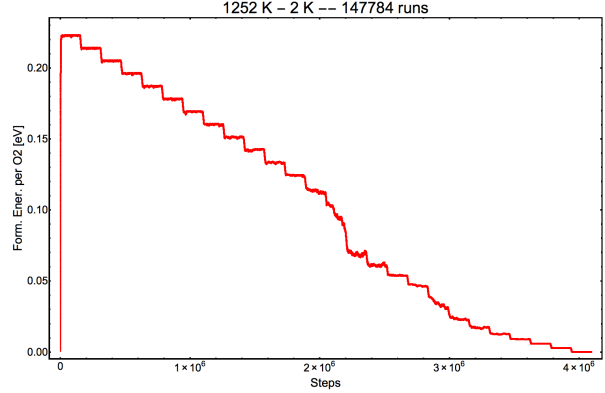

FIG. 4: Example of the annealing scheme used with the Metropolis Monte Carlo simulations. The system was initialized at high temperature, and then annealed down to just over absolute zero, in steps of 0.25 K. Every 50 K, the system was held at a constant temperature long enough to reach stability and take data for further analysis. Data taken at this 50 K intervals was used to produce the temperature-dependent plots of formation enthalpy and the entropic energy contribution

| Hubbard U | U3 | | U4 | | U5 | | U6 | | U7 | |
|---|---|---|---|---|---|---|---|---|---|---|
| Spin | s↑ | s↓ | s↑ | s↓ | s↑ | s↓ | s↑ | s↓ | s↑ | s↓ |
| AAAA-PPPP | 6.65 | 1.26 | 7.3 | 1.85 | 7.57 | 1.74 | 8.1 | 1.97 | 8.71 | 2.18 |
| AAAB-PPPP | 7.43 | 3.69 | 7.85 | 4.09 | 8.35 | 5.14 | 8.89 | 5.93 | 9.49 | 6.76 |
| AABB-PPPP | 7.03 | 3.96 | 7.37 | 4.64 | 7.78 | 5.33 | 8.24 | 6.07 | 8.76 | 6.68 |
| AABC-PPPP | 7.3 | 3.92 | 7.72 | 4.65 | 8.15 | 5.35 | * | * | 9.84 | 6.82 |
| ABCD-PPPP | 5.91 | 1.32 | 6.49 | 1.92 | 6.88 | 1.62 | 8.11 | 1.97 | 8.7 | 2.17 |
| AAAA-PPNN | 3.82 | 3.82 | 4.54 | 4.54 | 5.27 | 5.27 | 6.07 | 6.07 | 6.91 | 6.91 |
| AAAB-PPNN | 3.91 | 3.91 | 4.59 | 4.59 | 5.32 | 5.32 | 6.1 | 6.1 | 6.92 | 6.92 |
| AABB-PPNN | 4.13 | 4.13 | 4.83 | 4.83 | 5.57 | 5.57 | 6.35 | 6.35 | 7.21 | 7.21 |
| AABB-PNPN | 4.21 | 4.21 | 4.89 | 4.89 | 5.59 | 5.59 | 6.36 | 6.36 | 6.89 | 6.89 |
| AABC-PPNN | 4.39 | 4.13 | 5.1 | 5.1 | 5.84 | 5.55 | * | * | * | * |
| ABCD-PPNN | 3.94 | 3.94 | 4.66 | 4.66 | 5.42 | 5.42 | 6.23 | 6.23 | 7.12 | 7.12 |
| AABC-PNPN | * | * | * | * | * | * | * | * | * | * |

TABLE II: Table of bandgaps calculated with the hybrid functional HSE06. Antiferromagnetic structures show equivalently high bandgap in both spin channels, while ferromagnetic structures show a higher bandgap in the positive spin channel and a lower bandgap in the negative spin channel. Ferromagnetic structures of intermediate geometric disorder have a smaller disparity in spins than structures fully-ordered or fully-disordered.

\AtBeginShipout\AtBeginShipoutDiscard

\end{document}